\documentclass{emulateapj}
\def\stacksymbols #1#2#3#4{\def\theguybelow{#2}
        \def\verticalposition{\lower#3pt}
        \def\spacingwithinsymbol{\baselineskip0pt\lineskip#4pt}
        \mathrel{\mathpalette\intermediary#1}}
\def\intermediary #1#2{\verticalposition\vbox{\spacingwithinsymbol
        \everycr={}\tabskip0pt
        \halign{$\mathsurround0pt#1\hfil##\hfil$\crcr#2\crcr
                \theguybelow\crcr}}}
\def\lta{\stacksymbols{<}{\sim}{2.5}{.2}}
\def\gta{\stacksymbols{>}{\sim}{3}{.5}}

\begin{document}

\title{ENERGETICS OF X-RAY CAVITIES AND RADIO LOBES 
IN GALAXY CLUSTERS}

\author{
William G. Mathews\footnotemark[1] and 
Fabrizio Brighenti{\footnotemark[1]$^,$\footnotemark[2]}
}
%Andreas Faltenbacher\footnotemark[1],
%David A. Buote\footnotemark[3],
%Philip J. Humphrey\footnotemark[3],
%Fabio Gastaldello\footnotemark[3],
%\& Luca Zappacosta\footnotemark[3]

\footnotetext[1]{UCO/Lick Observatory,
Dept. of Astronomy and Astrophysics,
University of California, Santa Cruz, CA 95064}

\footnotetext[2]{Dipartimento di Astronomia,
Universit\`a di Bologna,
via Ranzani 1,
Bologna 40127, Italy}

%\footnotetext[3]{Dept. of Physics \& Astronomy, Univ. California 
%Irvine, 4129 Frederick Reines Hall, Irvine, CA 92697}

\begin{abstract}
We describe the formation and evolution of
X-ray cavities in the hot gas of galaxy clusters.
The cavities are formed only with relativistic
cosmic rays that eventually diffuse into
the surrounding gas.
We explore the evolution of cavities formed with
a wide range of diffusion rates but which are
otherwise similar.
In previous numerical simulations, in which
cavities are formed by injecting ultra-hot but
non-relativistic gas, cavity formation contributes 
thermal energy that may offset
radiative losses in the gas, thereby helping to solve the
cooling flow problem.
Contrary to these results, we find that X-ray
cavities formed solely from cosmic rays have
a global cooling effect.
Most cosmic rays in our cavity evolutions 
do not move beyond the cooling radius  
and, as the cluster gas is displaced, 
contribute to a global expansion of the cluster gas.
%and therefore contribute to a global cluster expansion for 
%times comparable to the cluster age.
%Successive cavities can pump up the 
%hot gas atmosphere, driving a significant mass outflow. 
%This global expansion accounts for most of the
%cooling associated with cavity formation.
As cosmic rays diffuse away from the cavities,
the nearby gas becomes buoyant, resulting in
a significant outward mass transfer within the cooling radius, 
carrying relatively 
%The outward circulation of buoyant gas transports relatively 
low entropy gas containing cosmic rays 
to outer regions in the cluster where 
it remains for times exceeding the local cooling time
in the hot gas. 
This post-cavity mass outflow due to cosmic ray buoyancy 
may contribute significantly 
toward solving the cooling flow problem. 
For example the mass inflow in the Virgo cluster due to 
radiative cooling can be balanced by buoyant outflow 
if only a fraction $\sim0.0005$ of the accretion energy 
onto the central black hole 
inflates X-ray cavities with cosmic rays. 
We describe the energetics, size, stability and buoyant
rise of X-ray cavities in detail, showing how
each depends on the rate of cosmic ray diffusion.
%$PV$ is an inaccurate measure of $E_{cav}$.
\end{abstract}


\section{Introduction}
The hot gas in galaxy clusters loses energy by 
X-ray emission but does not cool to low temperatures. 
In recent attempts to solve this cooling flow problem 
it has been almost exclusively assumed that feedback 
energy from accretion onto cluster-centered massive black holes 
can be delivered to 
the hot gas in a manner that maintains the observed temperature 
and density profiles in spite of radiation losses. 
The solution to this problem is not straightforward. 
Even when heat is supplied to the cluster gas 
in an {\it ad hoc}, optimized, ideally fine-tuned manner, 
either concentrated or distributed over many tens of kiloparsecs,
it is found that 
the temperature and density profiles deviate strongly 
from those observed (Brighenti \& Mathews 2002; 2003). 
For example, highly idealized flows in which radiative cooling 
is perfectly balanced by local heating at every radius 
are inconsistent with the secular increase in gas density 
associated with 
stellar mass loss in the cluster-centered galaxy. 
In most galaxy groups and clusters 
a positive temperature
gradient ($dT/dr > 0$) is observed
in the inner regions. 
Since the coolest gas 
is closest to the central source of AGN heating, 
this central gas must be heated with exquisite  
precision (on short timescales) to maintain the 
low temperatures observed. 

In spite of these difficulties, 
a variety of heating mechanisms continue 
to be investigated to understand the energetics of hot cluster gas. 
Heating is usually assumed to be 
associated with the formation of X-ray cavities 
by jet feedback energy proceeding from the central black hole. 
Weak shock waves emerge when 
the cluster gas is displaced as cavities form 
(see McNamara \& Nulsen 2007 for a review).
Weak shocks, observed in a few clusters 
(e.g. Perseus: Fabian et al 2003; Virgo: Forman et al. 2005; 2007), 
have the desirable capability of dissipating 
AGN feedback energy over 
large regions of the cluster gas,  
as emphasized by Fabian et al (2003). 
However, energy dissipation by outwardly propagating 
shock or sound waves 
is disproportionally 
concentrated in the central regions of clusters 
where the gas density gradient is smallest.  
Over time the cumulative wave dissipation in 
this central gas (through which all the 
waves must pass) causes its temperature to 
become hotter than observed in any cluster (Fujita \& Suzuki 2005; 
Mathews, Faltenbacher \& Brighenti 2006).
It may be possible to forestall cooling in a limited region 
of a particular cluster with a proper choice 
of shock Mach numbers (or sound wave frequencies 
and amplitudes), but this level of fine-tuning seems contrived.

X-ray cavities are thought to provide a
convenient measure of the amount of feedback power delivered
to the hot gas. 
Cavities are observed in $\sim20-25$\% of X-ray bright 
clusters (Birzan et al. 2004; Rafferty et al. 2006), 
but the incidence of cavities increases as the 
cooling time of the central gas decreases 
(Dunn \& Fabian 2006). 
The work done in displacing a volume $V$ of cluster gas at pressure
$P$ is $PV$ and the energy of the material inside the cavity
is $PV/(\gamma - 1)$.
This corresponds to a total cavity energy of
$E_{cav} = [\gamma/(\gamma-1)]PV = 4PV$
where $\gamma = 4/3$ is often chosen because many (young) cavities are
filled with radio synchrotron emission from relativistic
electrons and there is little or no observable evidence for 
hot thermal gas inside cavities (e.g. McNamara \& Nulsen 2007).

Assuming that the local buoyancy time $t_{buoy}$ 
in the cluster gas is a measure of 
the cavity lifetime, Rafferty et al. (2006) estimate the 
``cavity jet power'' 
\begin{displaymath}
P_{cav} = {E_{cav} \over t_{buoy}}.
\end{displaymath}
Rafferty et al. find that $P_{cav}$ exceeds the total X-ray luminosity 
$L_x$ within the cooling radius 
(defined where the cooling time is 7.7 Gyrs) 
in about 60\% of their sample clusters with cavities. 
They suggest that clusters with $P_{cav} < L_x$ -- and the 
many clusters with no known cavities -- are in a phase of their  
feedback energy cycle that is just now recovering 
from a recent heating episode.  
Rafferty et al. suggest that all the energy within the cavities 
$PV/(\gamma - 1)$ may be available to heat the cluster gas, 
not just the $PV$ work done by the expanding cavity. 
This implies that the (thermal) energy of material within the 
cavity is ultimately shared with the cluster gas. 
Furthermore, for those clusters in which the total cavity jet power
is favorable ($P_{cav} > L_x$), it is assumed that this energy is 
distributed throughout 
the cluster gas in a manner that preserves 
the characteristic gas temperature 
and density profiles observed in galaxy clusters. 
Implicit in this assumed energy distribution 
is the requirement that the cavity-forming 
energy is delivered to the cluster gas 
in an approximately isotropic manner relative to the 
central black hole. 

Another less often considered possible resolution to the cooling 
flow problem is the hypothesis that gas is only heated at or near the 
central black hole and then is buoyantly transported far out 
into the cluster gas, 
i.e. a circulation flow (Mathews et al. 2003; 2004).
One of the initial motivations for mass circulation was 
the two-temperature (and therefore buoyant) 
flow observed in the galaxy group NGC 5044 (Buote et al. 2003). 
However, Temi, Brighenti \& Mathews (2007a,b) have recently found 
far-infrared emission from 
dust extending out to 5-10 kpc around many group-centered, X-ray 
luminous elliptical galaxies. 
Since the dust lifetime to sputtering destruction 
in the hot gas is only 
$\sim10^7$ yrs, this observation provides additional strong 
evidence for ongoing buoyant outflow from dust-rich cores 
in the central galaxies. 
Additional support for buoyant outflow is provided by 
the large regions of iron enrichment in the hot gas 
surrounding cluster-centered elliptical galaxies 
(De Grandi et al. 2004). 
Since the 50-100 kpc size of these regions enriched by Type Ia 
supernovae greatly exceeds that of the central galaxy 
where they occur, 
an outward mass transfer is essential for their formation. 
Buoyant outflow is desirable since it can also  
preserve the observed gas density and temperature profiles 
(Mathews et al. 2004). 
Alternatively, 
these profiles can be preserved if mass circulates outward 
in momentum-driven, mass-carrying jets (Brighenti \& Mathews 2006).

In most theoretical studies of X-ray cavity evolution, 
it is assumed that the cavities are inflated with ultra-hot 
gas (occasionally but not always with $\gamma = 4/3$)
(e.g. Br\"uggen \& Kaiser 2002; 
Reynolds et al. 2005; Gardini 2007; Pavlovski et al. 2007).
If this heated gas is transported in jets from the 
central black hole, then this type of solution implicitly 
requires an outward mass circulation which in some simulations
can be quite large. 
For example, in the recent double jet 3D calculation of
Heinz et al. (2006), jets with power $W_j = 10^{46}$ ergs s$^{-1}$
enter the cluster gas from the origin with velocity
$v_j = 30,000$ km s$^{-1}$.
This translates into a mass flux of ${\dot M}_j = 2W/v^2 = 35$
$M_{\odot}$ yr$^{-1}$
and a total injected mass of
$M_j = 3.5 \times 10^9$ $M_{\odot}$ over their $10^8$ year
computation.
If ambient cluster gas is entrained in the jets, 
the outflowing mass could be increased further. 
%This mass outflow rate exceeds the traditional cooling flow rates 
%in most clusters: Perseus: 20 $M_{\odot}$ yr$^{-1}$, 
%Cygnus A: 31 $M_{\odot}$ yr$^{-1}$, 
%Hydra A: 16 $M_{\odot}$ yr$^{-1}$ etc., where we use 
%cooling flow estimates from Rafferty et al. (2006). 
%If such massive outflows are required to produce the desired 
%cavity heating, the cooling flow problem is solved 
%just with the outward mass transport required to offset cooling. 

The jet mass flux may be very much less if 
most of their energy density is in the form of cosmic rays.
While it is possible that jets do transport substantial 
masses of gas from the center
(as in Brighenti \& Mathews 2006), 
we explore here the energetics in the limiting case in which 
X-ray cavities are formed only with relativistic cosmic rays 
that can also diffuse into the cluster gas. 
For simplicity we assume that most of the cosmic rays in cavities 
arrive in jets and are not produced in local shocks. 
In this limit the cavities are formed with pure energy 
with no appreciable component of rest mass or momentum as in 
previous numerical simulations. 
Because of their diffusive nature, cosmic rays 
can eventually penetrate the cluster gas 
and contribute to the local pressure support. 
As a consequence the local cluster gas density is reduced by cosmic 
rays and becomes buoyant.  
Inhomogeneously distributed cosmic rays are a natural 
driver of buoyant mass outflow.
%The subsonic buoyant motion of regions with locally 
%enhanced cosmic ray pressure 
%can transport a substantial mass of cluster gas outward. 
Gas carried out by cosmic ray buoyancy
may never return to the cluster center.

However, we find that few cosmic rays move beyond the 
cooling radius,
% during a cluster cooling time, 
so the cluster gas expands and cools globally.
Although some heating is expected from shock waves, 
this heating is offset by the global 
cooling. 
Consequently, X-ray cavities containing cosmic rays 
result in a net cooling of the cluster gas, 
not heating as generally assumed. 
Buoyant mass outflow resulting from 
inhomogeneous cosmic rays and global expansion 
of the cluster gas 
may help to resolve the cooling flow puzzle. 

Although it is generally believed that X-ray cavities 
are formed with cosmic rays, this conjecture has 
only recently been tested with detailed calculations 
(Mathews \& Brighenti 2007a,b).
However, spherical steady state cluster flows including cosmic rays 
(e.g. B\"ohringer \& Morfill 1988;
Loewenstein et al. 1991)
have been developed to explore 
the possible large scale dynamical influence of 
cosmic ray pressure gradients and the dissipation
of their energy into the thermal gas. 

For simplicity in this initial treatment of 
the global time-dependent energetics of 
cluster cavity formation with cosmic rays, we do not 
specify the physical nature of the relativistic
fluid (electrons or protons) nor do we
calculate radio or inverse Compton fluxes/luminosities
which would require additional assumed parameters -- these 
important details will be considered in future papers. 
Nevertheless, the spatial distribution of cosmic rays 
we calculate  
defines the region where radio emission could be expected. 
Since we do not consider Coulomb heating,
the cluster gas responds adiabatically to
cosmic ray pressure gradients.

\section{Equations and Computational Procedure}

The combined Eulerian evolution of (relativistic) 
cosmic rays (CRs)
and thermal gas can be described with 
the following four equations:
\begin{equation}
{ \partial \rho \over \partial t} 
+ {\bf \nabla}\cdot\rho{\bf u} = 0
\end{equation}
\begin{equation}
\rho \left( { \partial {\bf u} \over \partial t}
+ ({\bf u \cdot \nabla}){\bf u}\right) = 
- {\bf \nabla}(P + P_c) - \rho {\bf g}
\end{equation}
\begin{equation}
{\partial e \over \partial t}
+ {\bf \nabla \cdot u}e = - P({\bf \nabla\cdot u}) 
%- (\rho/m_p)^2 \Lambda
\end{equation}
\begin{equation}
{\partial e_c \over \partial t}
+ {\bf \nabla \cdot u}e_c = - P_c({\bf \nabla\cdot u})
+ {\bf \nabla\cdot}(\kappa{\bf \nabla}e_c) 
+ {\dot S}_c
\end{equation}
%\begin{equation}
%{ \partial \rho z_{Fe} \over \partial t}
%+ {\bf \nabla}\cdot\rho z_{Fe} {\bf u} = 0
%\end{equation}
where we suppress artificial viscosity terms.
Pressures and thermal energy densities in the plasma
and cosmic rays are related respectively by
$P = (\gamma -1)e$ and $P_c = (\gamma_c - 1)e_c$
where we assume $\gamma  = 5/3$ and $\gamma_c = 4/3$.
The cosmic ray dynamics are described by
$e_c$, the integrated energy density over the cosmic ray
energy or momentum distribution,
$e_c \propto \int EN(E) dE \propto \int p^4 f(p)(1+p^2)^{-1/2} dp$.

The first three equations are the usual equations for 
conservation of mass, momentum and thermal energy in 
the hot thermal cluster gas. 
We do not include 
optically thin radiative losses since our intention here 
is to study the energetics of cavity 
creation alone without the complicating effects of 
a secular radiative energy loss and central cooling. 
Note that the CR pressure gradient in equation 2 
contributes to the motion of the thermal gas. 
This exchange of momentum between CRs and gas arises
as the CRs 
diffuse through magnetic irregularities (Alfven waves) 
that are nearly frozen into the hot thermal gas.
However, magnetic terms do not explicitly enter in 
the equations because typical magnetic fields in cluster 
gas $\sim 1-10\mu$G (Govoni \& Feretti 2004) 
are too small, i.e.  
the magnetic energy densities $\sim B^2/8\pi \lta 10^{-11}$
erg cm$^{-3}$ are generally much less than the thermal
energy density in the hot gas.
In addition, the Alfven velocity
$v_A = B/(4\pi \rho)^{1/2} = 2 n_e^{-1/2} B(\mu{\rm G})$
km s$^{-1}$ is typically much less than
the sound or flow speeds in cluster gas 
so the Alfven velocity of the magnetic scatterers can be 
ignored (e.g. Drury \& Falle 1986, Jones \& Kang, 1990).

Equation 4 above describes both the advection of CRs 
with the gas and their diffusion through the gas.
A mass conservation equation for the CRs is unnecessary
because of their negligible rest mass.
The CR diffusion coefficient $\kappa$ 
is difficult or impossible to calculate in the absence 
of detailed information about the magnetic field topology 
which is currently unknown. 
However, we expect that $\kappa$ may vary inversely with the 
density of the thermal gas, assuming that the magnetic 
field strength also scales with density.
For simplicity we ignore for now any dependence of $\kappa$ on 
CR particle momentum.
Since observed radio lobes are very approximately spherical,
we assume that $\kappa$ is isotropic, 
consistent with a highly irregular global magnetic field.
For these preliminary calculations it is not necessary
to specify the CR composition, either electrons or protons
can dominate as long as they are relativistic.
Finally, we assume that the total CR energy density is not 
substantially reduced by losses due to synchrotron emission 
or interactions with ambient photons or thermal particles 
during the cavity evolution time. 
Dissipation of cosmic ray energy into the thermal gas is 
probably not important in X-ray cavities since the gas 
temperature in the cavity rims is in fact relatively cooler.

The set of equations above are solved 
in ($r,z$) cylindrical coordinates using a 2D code 
very similar to ZEUS 2D (Stone \& Norman 1992).
To be specific, we study the evolution of X-ray cavities 
in the well-observed Virgo cluster 
using the analytic fits to the observed gas temperature 
and density profiles suggested by
Ghizzardi et al. (2004).
The computational grid consists of 100 equally spaced 
zones in both coordinates out to 50 kpc plus an additional 
100 zones in both coordinates that increase in size 
logarithmically out to $\sim1$ Mpc. 
We adopt reflection boundary conditions for gas and cosmic rays 
at the origin and, in view of the large computational domain,
also at the outer grid boundary.
We assume a spherical gravitational field 
${\bf g} = (g_z,g_r)$ that establishes exact initial hydrostatic 
equilibrium for the Virgo cluster gas pressure gradient.
The cosmic ray diffusion term in equation 4 is solved 
using implicit Crank-Nicolson differencing. 
While this differencing scheme is stable for all time steps,
we restrict each time step 
by the stability condition for explicitly-differenced 
diffusion as well 
as the Courant condition for numerical stability. 
Shocks are treated with a standard artificial viscosity.

We assume that the X-ray cavity is formed by 
CRs that propagate in a non-thermal jet 
from the central black hole 
(AGN) to some fixed radius.
The CRs are deposited in a gaussian-shaped 
sphere of characteristic 
radius $r_s = 2$ kpc located at 
${\bf r}_{cav} = (r,z) = (0,10~{\rm kpc})$, 
i.e. 10 kpc along the $z$-axis.
The CR source term in equation 4 is therefore
\begin{equation}
{\dot S}_c = {E_{cav} \over t_{cav}}
{e^{-(({\bf r - r}_{cav})/r_s)^2} \over \pi^{3/2} {r_s}^3} 
~~~{\rm erg}~{\rm  cm}^{-3}{\rm s}^{-1}
\end{equation}
when $t < t_{cav}$.
The integral of $ (r_s \pi^{1/2})^{-3}e^{-(r/r_s)^2}$ 
over space is unity.

In the calculations described below 
our principal objective is to explore the unique 
diffusive effects of cosmic rays on cavity formation 
and energetics. 
We therefore restrict the total CR energy of 
all calculations to 
$E_{cav} = 1 \times 10^{58}$ ergs and adopt 
$t_{cav} = 2 \times 10^7$ yrs as the CR injection time. 
We choose $t_{cav}$ to be consistent with X-ray cavity 
observations; it is shorter than the local buoyancy time 
in the cluster gas but sufficiently long not to produce 
strong shocks which are not commonly observed.
At times $t > t_{cav}$ when ${\dot S}_c = 0$, 
the total CR energy $E_{cr} = \int e_c dV$ over the grid 
volume remains very nearly constant 
but changes slightly due to advection in adiabatic 
compressions or rarefactions.

Since the CR diffusion coefficient is poorly known 
at present -- and may vary from one AGN event to another -- 
we consider a wide range of 
density-dependent coefficients, 
%page 128 of Latex book
\begin{displaymath}
\kappa = \left\{ 
\begin{array}
{r@{\quad:\quad}l}
10^{30} ~{\rm cm}^2 {\rm s}^{-1} & n_e \le n_{e0} ~{\rm cm}^{-3} \\
10^{30}(n_{e0}/n_e) ~{\rm cm}^2 {\rm s}^{-1} & n_e > n_{e0} ~{\rm cm}^{-3}
\end{array} \right. 
\end{displaymath}
In general, for reasons discussed above, 
we assume that the CR diffusion $\kappa$ varies 
inversely with cluster gas density. 
But $\kappa$ must be sufficiently large 
so that the CR density is approximately uniform in the X-ray cavities 
where the cluster gas density is lowest, 
i.e. we assume that CRs also diffuse inside the cavity.
For example when $E_{cav} = 1 \times 10^{58}$ ergs 
we find that the maximum cavity radius is $\sim 5$ kpc so 
we require that $\kappa$ in the cavities be at least 
$(5~\rm{kpc})^2/t_{cav} \approx 4 \times 10^{29}$ 
cm$^2$s$^{-1}$ and this condition is ensured by  
our adopted $\kappa(n_e)$ above.
In regions of higher $n_e$ 
the density parameter $n_{e0}$ determines the CR diffusion
coefficient.
In the following we consider 
$n_{e0} = 6 \times 10^{-3} - 6 \times 10^{-6}$ cm$^{-3}$.  
The largest CR diffusion coefficient $\kappa$
(corresponding to $n_{e0} = 6 \times 10^{-3}$)
is similar to that required 
by Mathews \& Brighenti (2007b) to explain a common age ($10^8$ yrs) 
for the large radio lobes and the cavity jet (thermal filament) 
observed by Forman et al. (2007) in the Virgo cluster. 

\section{Cavity Evolution with Different\\
Cosmic Ray Diffusivities}
%\centerline{COSMIC RAY DIFFUSIVITIES}
%\vskip.1in

Figure 1 shows the gas density $\rho(r,z)$ and cosmic ray 
energy density $e_c(r,z)$ at six times during the evolution 
of an X-ray cavity and its cosmic ray (radio) lobe. 
In Figure 1 $n_{e0} = 6\times 10^{-3}$ cm$^{-3}$ 
so the diffusion coefficient 
is rather large, but identical to that 
used by Mathews \& Brighenti (2007b) to describe the 
evolution of the cavity jet (thermal filament) in 
M87/Virgo.
The cavity in Figure 1 is formed 10 kpc along the (horizontal) 
$z$-axis during time $t_{cav} = 0.02$ Gyr. 
At time $0.024$ Gyr, shortly following $t_{cav}$, 
the cosmic rays (dotted contours) 
are tightly confined inside the cavity. 
By time $0.066$ Gyr the cosmic rays have diffused 
through the cavity walls forming a small radio lobe 
and the cavity has just disappeared 
-- we define the cavity as that region where the 
plasma density is lower than the original Virgo 
density by at least a factor 1/3.
A small vortex at $(r,z) = (5,20)$ kpc is visible at this time. 
During the first three times shown in Figure 1 
a (relatively cooler) 
thermal feature (the ``cavity jet'') is seen to 
rise along the $z$-axis. 
The evolution of this feature is described in detail 
in Mathews \& Brighenti (2007b) 
and we will not repeat that discussion again here.
During the three later times in Figure 1
the denser parts of the cavity jet have fallen back toward the 
center of Virgo and the hot gas dynamics 
become more quiescent.
However, the cosmic rays are seen to diffuse into 
a progressively larger region 
elongated along the $z$-axis.
Although not apparent in the gas density contours, 
there is a net outward buoyant migration of the 
gas in the region occupied by the cosmic rays 
(the radio lobe).

Figure 2 shows the cavity evolution when the cosmic ray 
diffusion is 1000 times lower 
(when $n_e > n_{e0} = 6\times 10^{-6}$ cm$^{-3}$) 
but with all other parameters unchanged. 
Now the cosmic rays are seen to be very tightly 
confined to the cavity region until at least 
time 0.1 Gyr when the cavity is still visible. 
In this evolution the outer parts of the cavity 
break away forming a vortex that migrates away from the $z$-axis, 
carrying its own cosmic rays. 
By time 0.3 Gyr the brightest parts of the 
radio image (corresponding to 
the largest cosmic ray energy density) should consist of two 
separate regions, a feature along the $z$-axis extending 
out to 40 kpc and the vortex at 
$(r,z) = (14,38)$ kpc. 
At this time (0.3 Gyr) we see an 
enhanced gas density that accompanies the cosmic rays 
along the $z$-axis. 
At still later times the cosmic rays continue to 
reside mostly along the $z$-axis but by time 0.9 Gyr a region of low 
cosmic ray energy density becomes visible 
in the $r$-direction 
along the trajectory of the receding vortex 
(now at $(r,z) = (25,28)$ kpc). 

When X-ray cavities are formed in an atmosphere initially 
at rest, as we assume here, the vortex region at time 
0.3 Gyr in Figure 2 would appear as a ring when viewed 
along the $z$-axis. 
To our knowledge no ring-shaped radio features have been 
observed. 
This could simply be due to the faintness of such regions 
since radio-synchrotron electrons tend to not to produce 
observable GHz emission after about $10^8$ yrs. 
Nevertheless, it is clear from Figures 1 and 2 that 
the radio lobe morphology can 
in principle provide valuable information concerning the 
cosmic ray diffusion coefficient about which very little 
is known at present. 
Finally, since gas phase metal abundances tend to increase toward 
the centers of clusters, regions of enhanced abundance 
are expected to accompany the outward buoyant migration of 
cosmic rays. 

As a further aid in interpreting Figures 1 and 2, 
Figure 3 shows the pressure profiles along the $z$-axis 
at four times for both evolutionary calculations. 
At early times the cavity 
is visible as the region where $P_c > P$. 
Within the cavity the pressure gradient 
($\sim dP_c/dr$) is nearly flat because of the enormous 
pressure scale height of the relativistic fluid. 
However, except near and within the cavities, 
the total pressure $P + P_c$ (dotted lines) 
deviates very little from the pressure profile in the 
cluster before the cavity was introduced 
(dot-dashed lines). 
This is a consequence of 
the largely subsonic character of the flows. 
After about 0.1 Gyrs in both calculations 
the cosmic rays diffuse sufficiently so that 
the pressure ratio $P_c/P \ll 0.2$, the current 
threshold of $\gamma$-ray detectability of cosmic ray protons 
if nearby clusters are completely filled with cosmic rays 
(Pfrommer \& Ensslin et al. 2004).
Although the cosmic ray pressure and energy 
density are negligible 
at these late times, we show below that 
cosmic rays continue to 
displace about the same volume of gas.
This results in a very long-lasting global expansion of 
cluster gas that can be seen by the small 
discrepancies in the $z$-axis gas pressure profile 
(solid lines) relative to that in the initial cluster
(dash-dotted lines) at times 0.3 Gyrs in Figure 3. 
Nevertheless, the close similarity of the initial and final gas pressure 
along the $z$-axis (and elsewhere) 
at late times is consistent with 
our finding that the gas density and temperature (and therefore  
entropy\footnotemark[1]) gradients are also only slightly 
affected by the cavity evolution. 
\footnotetext[1]{Figure 8 shows 
(for the evolution in Figure 2) an approximate  
$z$-axis gas entropy profile at 0.9 Gyrs 
relative to the initial atmosphere.
When plotted, $s(z)$ at 0.9 Gyrs is almost identical to that
in the original atmosphere from the 
origin to $z \approx 25$ kpc and beyond $z \approx 42$ kpc. 
In the intermediate region 
$25 < z < 42$ kpc the entropy at 0.9 Gyrs 
is depressed below the original profile 
approximately as shown in Figure 8.}

\section{Energetics and Evolution\\ of X-ray Cavities}

Figure 4 shows the evolution of 
cluster gas energies resulting from cavity-lobe 
formation with cosmic rays. 
The four panels show the evolution 
as the cosmic ray diffusion coefficient 
$\kappa$ decreases with $n_{e0}$ (in cm$^{-3}$) over a wide range:  
$n_{e0} = 6\times 10^{-3}$ (panel $a$), 
$6\times 10^{-4}$ (panel $b$),
$6\times 10^{-5}$ (panel $c$), and
$6\times 10^{-6}$ (panel $d$). 
The total cosmic ray energy integrated over the 
entire computational region $E_{cr} = \int e_c dV$ 
is shown with long dashed lines. 
$E_{cr}(t)$ is seen to rise until $t_{cav} = 2\times 10^7$ yrs 
as the cavities form,
then remain approximately constant after the cosmic ray 
source is turned off. 
$E_{cr}$ is not a strictly conserved energy. 
The small decrease in $E_{cr}(t)$ visible at times 
$t > t_{cav}$ can occur if the cosmic ray energy 
density is reduced by a secular 
advective expansion with the cluster gas.
This decrease in $E_{cr}(t)$ after $t_{cav}$ is 
stronger when $\kappa$ is smaller (panels $a \rightarrow d$)
since the cosmic rays are more confined near the 
$z$-axis where most of the gas expansion occurs. 
The (small) total kinetic energy in the cluster gas 
$E_{kin} = 0.5\int \rho (u_x^2 + u_y^2) dV$ 
is shown with dotted lines in each panel.
The change in potential energy compared to the 
initial Virgo atmosphere, 
$\Delta E_{pot} = \int \phi [\rho(t=0) - \rho] dV$ 
is shown with short dashed lines. 
The gravitational potential $\phi(R)$ is found 
from the initial M87/Virgo atmosphere by integrating 
$d\phi/dR = -(1/\rho) dP/dR$ 
where $R = (r^2 + z^2)^{1/2}$ is the radial coordinate.
The change in the total gas thermal energy 
relative to the original atmosphere 
$\Delta E_{th} = \int [e - e(t=0)] dV$ 
is shown with (the lower) solid lines in each panel.
Finally the dash-dotted line shows the total energy 
$E_{tot} = \Delta E_{th} + \Delta E_{pot} + E_{kin} + E_{cr}$ 
which is constant after time 
$t = t_{cav}$ to an excellent approximation and is 
equal to $E_{cav} = 10^{58}$ ergs as expected. 

Also shown in each panel of Figure 4 
is an approximate evaluation of 
the quantity $4PV$ where $V$ is the volume of the  
X-ray cavity at any time (arbitrarily defined as the 
sum over all grid zone volumes in which the gas density 
is less than $\rho(t=0)/3$) and $P$ is an estimate of 
the average pressure in the (sometimes non-contiguous)
zones containing the cavity.
When cosmic rays are strongly trapped within the cavity, 
we expect 
$4PV = E_{cav} = E_{tot}$.
In panel $a$ of Figure 4 we see that $4PV < E_{cav}$ 
which can be expected for larger $\kappa$ 
when cosmic rays diffuse through the cavity walls.
However, $4PV > E_{cav}$ is apparent in panels $c$ and $d$, 
although $4PV$ never
exceeds $E_{cav} = 10^{58}$ ergs by more than about 20\%.
This may be due to our rather 
approximate estimate of $P$ and $V$ or it may be a real 
inertial overshoot just after the cavity is formed. 
In any case, we include $4PV(t)$ 
in Figure 4 because this is the cluster gas heating energy 
proposed by Rafferty et al. (2006) and McNamara \& Nulsen
(2007) in their discussions of cavity energetics.

Figure 5 shows the approximate evolution of the cavity radius 
$r_{cav}(t)$ and its mean radius in the cluster $R_{cav}(t)$.
The cavity radius is found from the estimated cavity 
volume $V$ by assuming that the cavity is spherical,
$r_{cav} = (3V/4\pi)^{1/3}$.
The four lines for $r_{cav}$ and $R_{cav}$ in Figure 5
correspond to the four cosmic ray diffusivities 
$\kappa$ as it decreases  
in each panel in Figure 4: 
$a$, dotted line; 
$b$, short dashed line;
$c$, long dashed line;
$d$, solid line.
The buoyant trajectory of the cavities $R_{cav}(t)$ is 
independent of $\kappa$, but the cavities progress further 
along $R_{cav}(t)$ when $\kappa$ is smaller. 
The cavity lifetime is determined when $r_{cav} \rightarrow 0$.
Notice that the cavity lifetimes are similar for 
$n_{e0} = 6\times 10^{-5}$ (long dashed line), and
$n_{e0} = 6\times 10^{-6}$ (solid line), 
suggesting that the cavity evolution would not change 
much if $n_{e0}$ (and therefore $\kappa$) were reduced further.

The main result we wish to emphasize is the evolution of 
the cluster gas thermal energy $\Delta E_{th}(t)$. 
At early times during cavity formation when $t \lta t_{cav}$, 
$\Delta E_{th}$ increases because of 
heating due to the weak shock that propagates away from the 
expanding cavity. 
This initial shock heating increases as $\kappa$ decreases 
(panels $a \rightarrow d$ in Figure 4) 
because (for fixed $E_{cav}$ and $t_{cav}$) the shock 
strength increases when cosmic rays are more 
confined within the cavity. 
This is consistent with the 1D cavity evolution described 
by Mathews \& Brighenti (2007a).
However, after the cavity is formed 
($t \gta t_{cav}$) $\Delta E_{th}(t)$ 
decreases and becomes negative after 
$t \approx 3-5\times 10^7$ yrs. 
We see in Figure 4 that the final energy separation
between $\Delta E_{th}$ and $\Delta E_{pot}$
is independent of $\kappa$.
Also notice that in each panel of Figure 4
the average final value of the energy change
$0.5(\Delta E_{th} + \Delta E_{pot})$
is not zero, but is
approximately equal to the peak thermal energy
acquired from the shock, $\sim\Delta E_{th}(t_{cav})$. 
%Alternatively, after several $10^8$ yrs 
Although the total energy change in the cluster 
$\Delta E_{pot} + \Delta E_{th} + E_{kin}$ increases
as $\kappa$ decreases, 
the cluster gas experiences a net cooling 
($\Delta E_{th} < 0$)  
when cavities are formed with cosmic rays.
This global cooling is 
exactly opposite to the results discussed by other authors   
in which cavity formation is regarded as an important 
source of AGN heating required to balance radiative losses in the
cluster gas and reduce central cooling in cooling flows. 

The energy evolutions shown in Figure 4 are not qualitatively
altered when the total cavity energy $E_{cav}$ is increased.
For example when $E_{cav} = 10^{59}$ ergs
(deposited at $(r,z) = (0,10)$ kpc in time
$t_{cav} = 2\times 10^7$ yrs with $n_{e0} = 6\times 10^{-6}$
cm$^{-3}$) all energies are larger but the proportions
are similar to those in Figure 4.
Although the shock is stronger in this case, so is the
buoyant cooling and $\Delta E_{th}$ still becomes negative after
$\sim10^8$ yrs.
The maximum equivalent spherical radius of this
high energy cavity
is about $r_{cav} = 15$ kpc at $\log t \sim 8.2$,
but the cavity disappears soon afterward at $\log t = 8.4$
when $R_{cav} = 55$ kpc.
More energetic cavities are bigger and more buoyant,
but don't last proportionally longer.

\section{Global Expansion and Mass Outflow}

To verify that the cluster gas has in fact 
undergone a net expansion as a result 
of cavity formation, we computed the evolution of the 
cumulative gas mass distribution $M(R)$. 
First we sorted 
the computational zones in cylindrical coordinates 
in order of increasing spherical radius $R = (r^2 + z^2)^{1/2}$
then we integrated over the 7855 sorted zones within $R = 50$ kpc 
to determine $M(R)$. 
Figure 6 shows $M(R)$ in (one hemisphere of) 
the initial cluster ($t=0$) and 
after $t = 0.9$ Gyr for two limiting values of $\kappa$ 
($n_{e0} = 6\times 10^{-3}$ and $6\times 10^{-6}$ cm$^{-3}$).
To more clearly illustrate the evolution of $M(R)$ 
most of the $R$-variation in Figure 6 
has been removed by plotting $M(R)(R/10~{\rm kpc})^{-2}$. 
The noise in these plots of $M(R)$ arises because of the finite 
number of computational grid zones. 
Figure 6 clearly shows that mass has been removed from the inner 
regions of the cluster as a result of cavity formation, 
i.e. the cluster gas has experienced a net expansion. 
This expansion is remarkably insensitive to the 
cosmic ray diffusivity $\kappa$, particularly 
at $R \lta 25$ kpc. 
The evolution of $M(R)$ is expected to depend 
both on the cavity energy 
$E_{cav}$ and the location of the initial cavity in 
the cluster gas (10 kpc in our case).

The largest radius plotted in Figure 6, $R = 50$ kpc, 
is essentially the cooling radius for the cluster, 
i.e. the radius where the local cooling time is 7.4 Gyrs, 
which is comparable to the cluster age.
From this plot we estimate that 
$\Delta M_{10} = 1.14\times 10^8$ $M_{\odot}$ 
has been removed from within 
10 kpc for both values of $n_{e0}$.
At $R = 50$ kpc the mass outflow is 
$\Delta M_{50} = 3.6\times 10^8$ $M_{\odot}$ 
(for $n_{e0} = 6\times 10^{-3}$) 
and $\Delta M_{50} = 5.6\times 10^8$ $M_{\odot}$
(for $n_{e0} = 6\times 10^{-6}$). 
Values of $\Delta M$ 
refer to the total cluster mass flow in both hemispheres, 
assuming that a pair of cavities are produced 
by symmetric double jets. 

%According to Rafferty et al (2006), 
%the cooling rate in M87/Virgo is 
%${\dot M}_{cf} = 1.2 M_{\odot}{\rm yr}^{-1}$.
%Our own estimate u
Using the Ghizzardi et al. (2004) 
density and temperature profiles for Virgo, 
we estimate a feedback-free gas cooling rate of 
${\dot M}_{cf} \approx 8 M_{\odot}/{\rm yr}$.  
Therefore if cavities similar to the one we calculate here 
are formed with cosmic rays every $6\times 10^7$ yrs, 
this could result in a mass outflow at $R = 50$ kpc equal to the 
cooling inflow at $R = 0$ from radiation losses. 
%However, in that case the cavity lifetimes 
%taken from Figure 5 would indicate that a 
%cavity with $r_{cav} \sim 2-5$ kpc 
%should be visible in Virgo at all times, 
%but no such cavity is visible at present. 
%This could be explained by the difficulty of observing 
%cavities at larger $R$, particularly when they 
%are not in the plane of the sky 
%(e.g. Ensslin \& Heinz 2002; 
%McNamara \& Nulsen 2007). 
%The lack of visible large cavities in Virgo 
%can also be understood if the 
%mean cavity energy $E_{cav}$ is larger. 
%We have already 
%shown that an $E_{cav} = 10^{59}$ erg cavity 
%(with $n_{e0} = 6\times 10^{-3}$ cm$^{-3}$) 
%in Virgo can simultaneously explain the 
%observed length 
%of the thermal cavity jet and the large radius of the 
%outer radio lobe, both of which have the same age, $\sim 10^8$ yrs
%(Mathews \& Brighenti 2007b).
When $E_{cav} = 10^{59}$ ergs the total mass flowing
beyond 50 kpc after time 0.9 Gyr is much larger, 
$\Delta M_{50} = 4.6\times 10^9$ $M_{\odot}$. 
(Note that the buoyant mass outflow scales with $E_{cav}$, 
but the cavity lifetime does not.) 
Successive cavities in Virgo with $E_{cav} = 10^{59}$ erg 
appearing every $6\times 10^8$ years could 
balance the mass flow from radiative cooling  
${\dot M}_{cf} \approx 8 M_{\odot}/{\rm yr}$  
and also be consistent with the current absence of 
large cavities in M87/Virgo. 
Of course an expansion outflow at the cooling radius 
comparable to the steady state cooling rate  
${\dot M}_{cf}$ does not in itself 
shut down radiative cooling near the cluster center, 
but the cooling radius will become larger and the cooling 
time at every radius will increase due to the slightly 
lower gas density resulting from the global expansion.

\section{What Increases the Potential Energy\\
and Drives the Mass Outflow?}

The post-cavity mass flow has two aspects:
(1) an increase in $\Delta E_{pot}$ in Figure 4
and non-zero $\Delta M_{50}$ in Figure 6 that result 
from the global lifting that occurs when a new 
cavity is formed, and 
(2) a mass circulation driven near the $z$-axis 
by buoyant gas containing 
cosmic rays. We discuss each in turn.

\subsection{Global Lifting}

Although gas certainly 
flows beyond $R = 50$ kpc in all our cavity evolutions, 
the fraction of total cosmic ray energy within this radius 
$E_{cr,50}/E_{cr}$ varies. 
For example, at $t = 0.9$ Gyr and large $\kappa$
($n_{e0} = 6\times 10^{-3}$ cm$^{-3}$),
the cosmic ray energy in $R < 50$ kpc, 
$E_{cr,50} = 66.3\times 10^{56}$ ergs,  
is less than the total cosmic ray 
energy on the computational grid, 
$E_{cr} = 81.5\times 10^{56}$ ergs. 
In this case
about 20\% of the cosmic rays have moved beyond 50 kpc.
For smaller $\kappa$ ($n_{e0} = 6\times 10^{-3}$ cm$^{-6}$)
after this same time we find that
$E_{cr,50} = 41.8\times 10^{56}$ ergs  
is identical to the total energy 
$E_{cr}$, i.e. almost all cosmic rays are confined within 
50 kpc. 
(Both $E_{cr,50}$ and $E_{cr}$ refer to a single hemisphere.)
However, the mass $\Delta M_{50}$ flowing across $R = 50$ kpc 
is positive for both cases involving 
large and small diffusivities $\kappa$. 

Evidently mass can flow across the cooling radius 
even in the absence of local buoyant transport. 
This must arise because of the quasi-adiabatic 
expansion of the entire cluster gas when cavities are formed.
To understand this better, we begin by showing that 
the volume of gas displaced by cosmic rays $V_c$ depends on the 
local pressure $P_a$ in the cluster atmosphere and the total 
energy $E_{cr}$ of the cosmic rays, but not on the ratio 
of cosmic ray to gas partial pressures, 
$P_c/P$, which changes as the cosmic rays diffuse into the gas.
Imagine a uniform gas containing 
a spherical region of volume $V$ that contains 
a total cosmic ray energy $E_{cr}$. 
In pressure equilibrium $P_a = P_c + P$ where
$P_c = (\gamma_c -1)E_{cr}/V$ and therefore
$V = (\gamma_c - 1)E_{cr}/P_c$.
But the partial volume $V_c$ containing gas 
that is actually displaced by the cosmic rays is 
$V_c = (P_c/P_a)V = (\gamma_c - 1)E_{cr}/P_a$. 
Therefore as cosmic rays diffuse into larger volumes $V$, 
the total volume of gas that is displaced $V_c$ remains 
unchanged and independent of $\kappa$.
In a non-uniform cluster environment the volume of gas displaced by 
a given total energy of cosmic rays $E_{cr}$ varies with 
cluster radius as $P_a(R)^{-1}$.

$\Delta M_{50}$ can be estimated by assuming that all  
cosmic rays in the original cavity do not diffuse 
beyond their original cavity volume $V_{cav}$ 
and remain at their initial cluster radius $R_{cav} = 10$ kpc. 
In this limit, and after several cluster sound crossing times, 
we expect that the mass flow across $R = 50$ kpc 
will be equal to 
the gas mass displaced by the original cavity, 
$\Delta M_{50} \approx V_{cav} \rho(R_{cav}) = 
6.2\times 10^8$ $M_{\odot}$ where 
$V_{cav} = (4 \pi /3)r_{cav}^3$, 
$r_{cav} \approx 6$ kpc and  
$\rho(R_{cav}) = 4.72\times 10^{-26}$ gm s$^{-1}$ is the 
original cluster density at $R_{cav}$. 
This simple estimate of the outflowing mass is very similar to that 
found from Figure 6, 
$\Delta M_{50} = 5.6\times 10^8$ $M_{\odot}$.

However, discrepancies between computed and estimated masses
$\Delta M_{50}$ are expected because of several additional details.
The cosmic ray energy inside the estimated cavity volume is 
$E_{cr} = 3P_{10}V_{cav} = 101\times 10^{56}$ ergs,
where $P_{10} = P_a(R=10~{\rm kpc}) = 1.28\times 10^{-10}$ dyne cm$^{-2}$. 
However, this energy is about twice as large as the value of
$E_{cr}(t_{cav})$ in the low-$\kappa$ evolution
shown in panel $d$ of Figure 4, 
so the estimated value of $\Delta M_{50}$ above 
should be reduced by a factor of $\sim2$.
However, $\Delta M_{50}$ is expected to differ from 
$V_{cav} \rho(R_{cav})$ because of changes in the mean
volume $V_c$ of gas displaced by cosmic rays
after their outward motion during time 0.9 Gyr.
To estimate this suppose that the $e_c$-weighted mean
radius of cosmic rays at time 0.9 Gyr is $R = 30$ kpc 
(see Figure 2) where the gas pressure is 
$P_{30} = 0.608\times 10^{-10}$ dyne cm$^{-2}$. 
For constant total cosmic ray energy we expect 
the volume of gas displaced to vary as $V_c \propto {P_a}^{-1}$. 
Therefore if most of the cosmic rays are transported 
to $\sim30$ kpc, our original estimate of $\Delta M_{50}$ must 
be increased by $\sim P_{10}/P_{30} = 2.1$.
The combination of both corrections leaves our original 
estimate nearly unchanged.
Considering the uncertainties in these estimates, 
we conclude that the total mass flow past $R = 50$ kpc found
from Figure 6 for the low-$\kappa$ evolution,
$\Delta M_{50} = 5.6\times 10^8$ $M_{\odot}$, 
is consistent with a quasi-adiabatic expansion of the
Virgo atmosphere past radius 50 kpc due to gas displaced 
by cosmic rays when the cavity formed and 
no buoyant mass flow across $R = 50$ kpc is required.

\subsection{Buoyant Gas Circulation}

In addition to this atmospheric lifting,
buoyant mass flow largely within the cooling radius 
also contributes to
the post-cavity cluster evolution. 
Buoyant gas increases $\Delta E_{pot}$ and as it
expands we expect $\Delta E_{th}$ to decrease.
But every buoyant element must be accompanied elsewhere 
in the hot gas atmosphere by a downward replacement flow, 
largely balancing the buoyant changes in $\Delta E_{pot}$ and 
$\Delta E_{th}$. 
If cavities are created by feedback following a small accretion onto the 
central AGN, a buoyant mass circulation flow is a natural response to the 
inflow expected from radiative cooling. 
The low-$\kappa$ cavity evolution
($n_{e0} = 6\times 10^{-6}$ cm$^{-3}$) in Figure 2 
is particularly interesting because all buoyant motion
occurs within the cooling radius at 50 kpc.

%***must show that the mean $e_c$-weighted $R$ increases for this 
%solution -- but this could in principle result only from diffusion!
%
%What is missing is foolproof evidence for mass outflow due to 
%cosmic ray buoyancy!!! And that this gas  never returns  to 10 kpc.

Figure 7 
shows contours of the radial velocity $v_r(r,z)$ at four times 
during the low-$\kappa$ cavity evolution in Figure 2 
($n_{e0} = 6\times 10^{-6}$ cm$^{-3}$). 
(A plot of the radial velocity in 
the high-$\kappa$ cavity evolution shown 
in Figure 1 is almost identical.)
We plot $v_r(r,z)$ only near the $z$-axis since the radial 
velocities elsewhere in the atmosphere are very much smaller. 
Figure 7 shows a strong and sustained subsonic outflow 
until at least 0.1 Gyr. 
At $t = 0.3$ Gyr a backflow occurs near the $z$-axis 
as the densest parts of the thermal ``cavity jet'' fall back 
toward the cluster center (see Mathews \& Brighenti 2007b for 
more details). 
Later at $t = 0.5$ Gyr a mixture of positive and negative 
radial velocities is visible but all velocities are smaller 
and this trend toward quiescence continues until $t = 1$ Gyr. 

Because the entropy decreases monotonically with radius in the initial cluster 
atmosphere, we can use the entropy as a tracer to confirm that 
there has been a net mass outflow within 50 kpc 
during the low-$\kappa$ evolution (which is essentially adiabatic 
apart from the weak shock wave).
We define the gas entropy as $s = 10^{-33}(e/\rho^{5/3})$ in 
cgs units. 
In addition we expect that the buoyancy and outward flow  
of each gas element will increase with the partial pressure 
of the cosmic rays, $P_c/(P + P_c)$.
In the upper panel of 
Figure 8 we plot $P_c/(P + P_c)$ against 
$\Delta s = s(t = 0.9) - s_{atm}$, 
the change between the entropy $s$ at 0.9 Gyr 
and the entropy at the same location in the original 
Virgo atmosphere, $s_{atm}$.
Each point corresponds to a computational grid zone.
In those regions where $P_c/(P + P_c)$ is largest 
we see that $\Delta s$ decreases systematically with $P_c/(P + P_c)$, 
indicating that gas containing more cosmic rays at time 0.9 Gyr 
originally came from high entropy regions closer to the cluster center.

The lower panel of Figure 8 shows (with small points) the 
location in the cluster at time 0.9 Gyr that contains gas 
with $P_c/(P + P_c) > 0.3$ and the open squares mark those 
(significantly overlapping) zones where $\Delta s < -0.1$ 
for the low-$\kappa$ cavity evolution. 
Gas with the highest cosmic ray content is also the gas with the 
lowest entropy compared to that at the same radius 
in the original cluster. 
The spatial distribution of this low entropy gas is similar 
to the cosmic ray contours in the final panel of Figure 2. 
%In regions of Figure 8 where $\Delta s < 0$ the entropy gradient 
%$ds/dR$ at first nearly 
%vanishes, then, with increasing cluster radius $R$, 
%$s$ rises back to the 
%initial cluster profile $s_{atm}(R)$ -- i.e. in equilibrium 
%$ds/dR \ge 0$. 
Low entropy regions in the lower panel of Figure 8 
also have a slightly smaller gas density than adjacent gas 
at the same cluster radius $R$ without cosmic rays. 
During times of interest for the cluster evolution, the gas in 
these post-buoyant regions containing cosmic rays 
will never return to the cluster core where it initially resided. 
Figure 8 shows conclusively that low entropy
gas originally at small cluster 
radii has been buoyantly transported outward during the
low-$\kappa$ post-cavity evolution, but very little has moved beyond
50 kpc by time 0.9 Gyr. 
Cosmic rays from cavities are driving a mass circulation 
in the cluster gas out to a large fraction of the cooling radius.

Evidently the long-term 
decrease in cluster gas mass $\Delta M_{10} = 1.14\times 10^8$ 
$M_{\odot}$ within $R = 10$ kpc where the cavities formed 
arises due to the combined effects of the jet of thermal 
gas along the $z$-axis and the cosmic ray buoyancy that 
prevents much of this gas from returning back to $R = 10$ kpc. 
Recall that $\Delta M_{10}$ is nearly independent of the 
cosmic ray diffusion coefficient. 
The cosmic ray energy required to remove a solar mass 
of gas from $R = 10$ kpc is about 
$e_{\odot} = E_{cav}/\Delta M_{10} = 8.8\times 10^{49}$
erg ${M_{\odot}}^{-1}$. 
If ${\dot M}\delta t = 8$ $M_{\odot}$ 
would cool each year in Virgo due to radiation 
losses alone, the cosmic ray luminosity required 
for a buoyant outflow to balance ${\dot M}$ is 
$L_{cr} = e_{\odot} {\dot M} = 7.0\times 10^{50}$ erg yr$^{-1}$ 
or $2.2\times 10^{43}$ erg s$^{-1}$.
However, the rate that energy is created by accretion 
%in the absence of feedback the accretion energy 
onto the $\sim3 \times 10^9$ $M_{\odot}$ central black hole 
in M87/Virgo is 
$L_{acc} \approx 0.1{\dot M} c^2 = 4.6\times 10^{46}$ 
erg s$^{-1}$. 
The cosmic ray luminosity $L_{cr}$ required to 
balance a radiating cooling mass inflow in Virgo is $\sim 2000$ times 
less than the accretion luminosity generated by the black hole 
$L_{acc}$. 
Therefore, if only a fraction $\epsilon_{cr} \approx 5\times 10^{-4}$ of 
the accretion energy converts to cosmic rays that inflate 
X-ray cavities, this might be sufficient to shut down 
the cooling flow. 
A subsequent calculation with multiple cavities and 
radiative cooling will be required to verify this.

\section{Discussion}

In previous numerical simulations of X-ray cavity evolution 
(e.g. Br\"uggen \& Kaiser 2002;
Reynolds et al. 2005; Gardini 2007; Pavlovski et al. 2007) 
as well as in the recent review of 
McNamara \& Nulsen (2007)
it has been concluded that the cavities have 
an important net heating effect on the cluster gas.
This result can be understood because 
the simulated cavities were inflated with ultra-hot 
(but non-relativistic) gas. 
As these cavities are initially inflated,  
they perform an amount of heating $\sim PV$ 
on the surrounding cluster gas, 
but the thermal energy inside the cavities also 
contributes to the cluster thermal energy budget.
When we perform calculations in which cavities
are inflated with ultra-hot gas,
we also find a (small) net heating 
when the total cluster thermal
energy includes the energy injected inside the cavity.
In this case the total energy introduced by the cavity is
$ [\gamma/(\gamma-1)]PV = (2.5-4)PV$, depending on the
assumed value of $\gamma$ inside the cavity. 
(After times $\gta 10^8$ yr most of the energy of hot-gas cavities 
is stored in the potential energy of the cluster.) 
The approximate values of $4PV$ plotted in Figure 4
show the estimated cavity heating when 
cavities are formed with ultra-hot gas 
(e.g. Birzan et al. 2004; Rafferty et al. 2007;
McNamara \& Nulsen 2007).

But ultra-hot gas is not an appropriate 
substitute for cosmic rays.
When the diffusion $\kappa$ is sufficiently small 
and cosmic rays are strongly trapped within the 
cavities, we find here that the expanding cavities 
also heat the gas by $\sim PV$. 
This can be seen in Figure 4 
where $\Delta E_{th}(t_{cav}) \approx PV$. 
%However, when cosmic rays ultimately diffuse through
%the cavity walls, 
%a negative work comparable to $-PV$ is performed on the
%cluster gas and 
%it is possible that this is a (small) part
%of the global post-cavity 
%cooling ($\Delta E_{th} < 0$) seen in Figure 4.
But the cosmic ray contents of our cavities never 
contribute to the cluster thermal energy.
Instead, the displacement of cluster gas by the 
diffusing, buoyant cosmic rays results in an 
overall expansion and cooling of the cluster gas.

It should also be recognized that the introduction of 
ultra-hot gas in previous numerical 
simulations of cavity evolution 
implicitly suggests that mass as well as 
cosmic ray energy 
is transported out from the central AGN, 
i.e. this conventional means of 
cavity formation must be regarded 
as a (mass) circulation flow 
similar to those we have considered 
(Mathews et al. 2003; 2004). 
This assumed non-relativistic mass outflow can be quite large as 
we discussed in the Introduction. 
While we assume here that the AGN produces 
jets of pure relativistic energy, 
it is possible that real jets do in fact carry 
non-relativistic mass away from the central AGN 
(as in Brighenti \& Mathews 2006).
Nevertheless, current X-ray observations 
have been unable to detect thermal emission from 
ultra-hot gas in the cavities 
(e.g. McNamara \& Nulsen 2007, but see 
Mazzotta et al. 2002). 

\section{Conclusions}

It is generally agreed that the dominant source 
of feedback energy in galaxy clusters is
accretion onto central black holes (AGNs).
Some of this accretion energy is thought to be transported
out into the cluster gas by jets, forming X-ray cavities.
In previous numerical simulations of X-ray cavity
evolution the cavities were formed by introducing
ultra-hot, non-relativistic gas at some radius in
the cluster gas.

For the first time we consider cavities that are formed
exclusively by cosmic rays that can both diffuse through
the cluster gas and be advected by it.
We discuss the energetics 
of X-ray cavity and radio lobe evolution in the hot 
gas within the Virgo galaxy cluster. 
For simplicity we consider cavities that are formed 
in time $t_{cav} = 2\times 10^7$ yrs by cosmic rays 
with total energy $E_{cav} = 10^{58}$ ergs 
but which have a wide range of  
cosmic ray diffusion coefficients $\kappa$ 
since this is currently 
the most uncertain parameter in cavity-lobe evolution.
For simplicity, we do not consider here secular 
energy losses in the cosmic rays due to synchrotron 
emission, Coulomb heating, inverse Compton losses, 
spallation etc., so the total cosmic ray 
energy can only change due to local expansion or 
compression by the thermal gas. 
In this approximation cosmic ray pressure gradients
interact with the cluster gas in a perfectly adiabatic fashion.
  
We conclude that:
 
\vskip.1in\noindent
(1) As cavities expand during formation, 
they generate weak shock waves that propagate 
into the surrounding cluster gas 
where the shock energy is dissipated.
When the cosmic ray diffusion $\kappa$ is low, 
the maximum shock heating $\sim PV \sim E_{cav}/4$ is approached.
As $\kappa$ increases the cosmic ray pressure 
gradients are lower and the shock heating is considerably less. 
In general, $4PV$, which varies with time, 
is an inaccurate measure of $E_{cav}$.  
These results are consistent with our earlier 
studies of 1D cavity evolution in a uniform 
gas (Mathews \& Brighenti 2007a).

\vskip.1in\noindent
(2) The longevity of visible X-ray cavities and
the radial distance that they move out
in the cluster gas during their lifetimes
both increase as the cosmic
ray diffusion $\kappa$ decreases. 
Unlike cavities formed with hot gas 
(e.g. Gardini 2007), cosmic ray cavities remain coherent
for $\gta10^8$ yrs.

\vskip.1in\noindent
(3) In spite of shock heating, 
cavities formed by cosmic rays 
have a net cooling effect on the cluster gas. 
This is unlike the cluster heating when 
cavities are inflated largely with ultra-hot 
but non-relativistic gas, as commonly assumed, 
that later contributes to the cluster thermal energy.

\vskip.1in\noindent
(4) The qualitative character of post-cavity 
energetics remains similar when the 
cosmic ray diffusion coefficient $\kappa$ 
varies over a wide range. 
After several $10^8$ yrs the total energy change in the 
cluster $\Delta E_{pot} + \Delta E_{th} + E_{kin}$ increases 
as $\kappa$ decreases.

\vskip.1in\noindent
(5) As cosmic rays displace cluster gas, 
the entire cluster atmosphere expands 
and cools, increasing the cluster potential energy and 
decreasing its thermal energy.
Successive cavities will result in an irreversible 
expansion of the cluster gas as long as most of the cosmic rays 
remain trapped within or near the cooling radius, 
as in our computed post-cavity evolutions. 
%For the cluster we consider this cavity-generated
%mass outflow is comparable
%to the inward mass flow in a traditional cooling
%flow driven by radiation losses. 

\vskip.1in\noindent
(6) Inhomogeneously distributed 
cosmic rays are an important source of buoyancy 
and mass flow in the cluster gas.
As the cluster gas surrounding cavities is 
penetrated by diffusing cosmic rays, 
this gas rises in the cluster potential, 
driving a net mass outflow in the cluster gas. 
Radial mass circulation occurs 
even when no gas flows out in the AGN jets. 

\vskip.1in\noindent
(7) After buoyant gas containing cosmic rays 
flows from the cluster core into distant, 
low-density regions, it remains there never to return 
during times comparable to the cluster age. 
Because of the necessarily positive entropy gradient 
in the pre-cavity cluster,
the entropy in the post-buoyant gas in these distant 
regions is slightly lower
than elsewhere at the same cluster radius.

\vskip.1in\noindent
(8) To quantify the previous conclusion, 
suppose a mass $\Delta M_{R}$ within the cluster radius $R$ 
where the X-ray cavity forms is irrevocably transported 
by cosmic ray buoyancy to distant regions in the cluster gas.
We find that a cosmic ray energy of 
$e_{\odot} = E_{cav}/\Delta M_{R} \approx 10^{50}$
ergs removes a solar mass of gas from the cavity site. 
For example in the Virgo cluster a cosmic ray luminosity of only 
$L_{cr} = e_{\odot} {\dot M} = 2\times 10^{43}$ erg s$^{-1}$ 
can remove gas at the cooling flow rate 
${\dot M} \approx 8$ $M_{\odot}$ yr$^{-1}$ that would 
otherwise occur in the absence of feedback energy. 
Only a fraction $\epsilon_{cr} \approx 5\times 10^{-4}$
of the accretion energy onto the central black hole
in M87/Virgo, 
$L_{acc} \approx 0.1{\dot M} c^2$, 
if converted to cosmic rays that inflate
X-ray cavities, is in principle 
sufficient to shut down the cooling flow.

\vskip.1in

%Current radio and X-ray observations are 
%consistent with cavities filled only with relativistic 
%gas, but it is possible that some non-relativistic 
%gas is also transported in jets from the central AGN, 
%although we do not consider that case here.

Our conclusions suggest that cosmic ray buoyancy 
provides an important new means of 
understanding or possibly solving the cooling flow problem 
with an outward mass circulation that may balance the inexorable  
inflow due to radiation losses. 
This outward mass flow due to cosmic ray buoyancy has the 
added attraction of not heating the gas too much -- in fact 
cavity formation with cosmic rays produces a net cooling. 
%A slow mass outflow along and near the cavity jet axis 
%may preserve the positive gas temperature gradient 
%within $\sim 0.1r_{virial}$ that is commonly observed.

It remains to be determined if the radio emission
or $\pi^0$ gamma emission from galaxy clusters are at
levels that are consistent with the buoyant activity
required to arrest the cooling flow rate.  
After a few $10^8$ years most of the cosmic rays 
will be in very low density gas where such emission will 
be greatly reduced.

%We plan to expand our study of cosmic ray 
%cavities to study the evolution and energetics 
%in the presence of radiative cooling and 
%when other parameters are varied, e.g. $E_{cav}$, 
%$t_{cav}$. 
%However, we have already described in detail the 
%evolution of cavities with $E_{cav} = 5-10\times 10^{58}$ ergs 
%(and $t_{cav} = 2\times 10^7$ yrs) in the M87/Virgo 
%cluster.
%By comparison of the age of the $\sim30$ kpc long thermal 
%cavity jet with that of the outer radio lobe in Virgo, 
%we argued that both of these features were formed in 
%the same cavity-forming event about $10^8$ yrs ago 
%(Mathews \& Brighenti 2007b). 
%We achieved this agreement with rather low cosmic ray 
%diffusion coefficients $\kappa$ similar to that in panel $a$ of 
%Figure 4. 
%When $\kappa$ is very small, 
%the cosmic rays remain confined along the 
%direction of the buoyant trajectory of the cavity. 
%It is unclear if the low-$\kappa$ cosmic ray geometries 
%illustrated in Figure 2 actually occur in nature. 
%In principle the shapes and sizes of radio lobes in galaxy 
%clusters can reveal much about $\kappa(n_e)$ and its 
%variation with particle energy. 

%kappa in cavity walls with B perp

\vskip0.1in
Studies of the evolution of hot cluster gas 
at UC Santa Cruz are supported by
NASA and NSF grants for which we are very grateful.

%\end{document}

\clearpage
\begin{figure}%1
\centering
\vskip.5in
\hspace{2.in}
\includegraphics[bb=62   5 350 437,scale=1.0,angle= 0]
{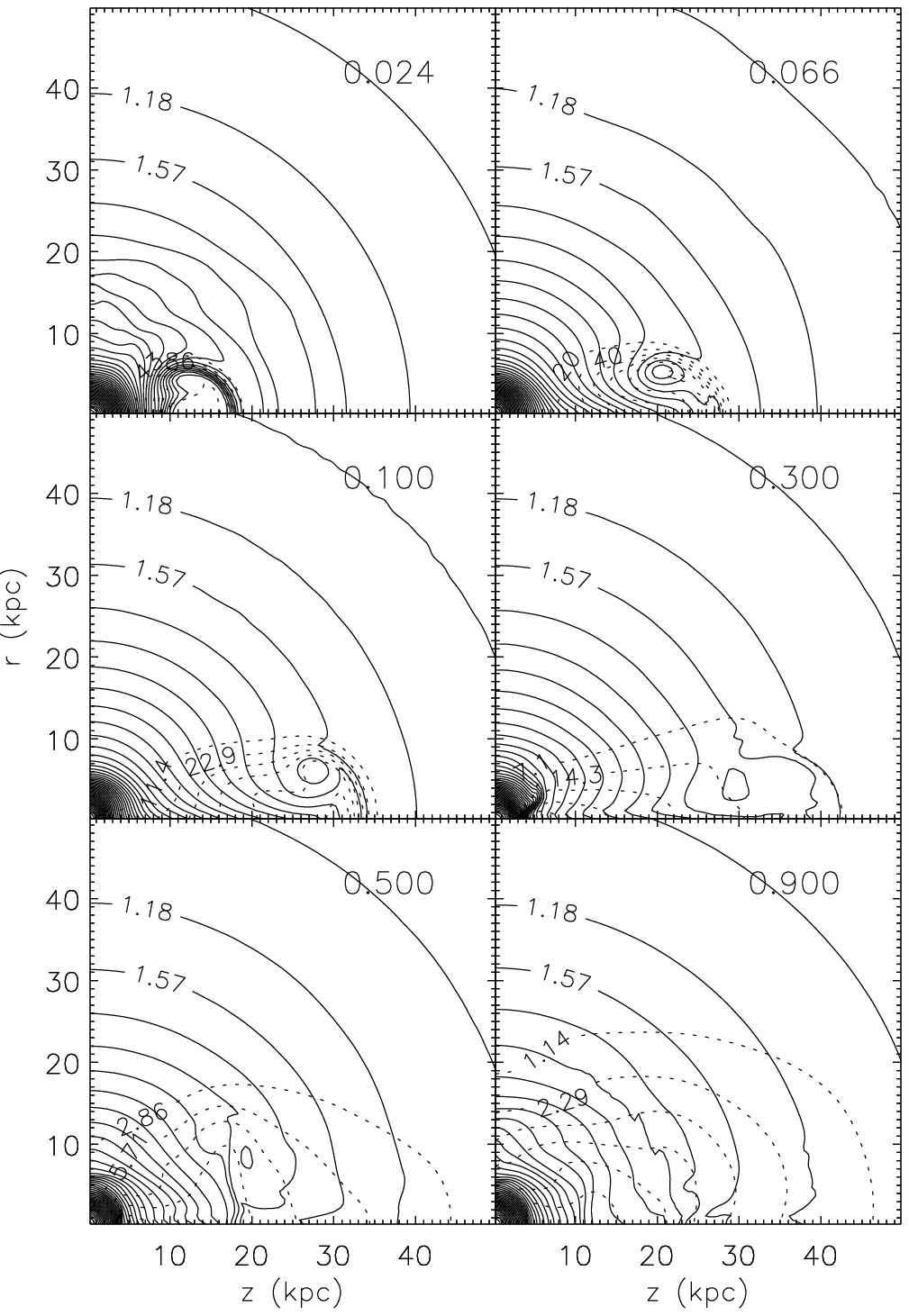}
\vskip.5in
\caption{
Evolution of an X-ray cavity in Virgo with cosmic ray diffusion 
parameter $n_{e0} = 6\times 10^{-3}$ cm$^{-3}$ 
at six times shown in Gyrs at the upper right of each panel.
Solid lines show the gas density contours $\rho(r,z)$ in units of 
$10^{-26}$ g cm$^{-3}$. 
Dotted lines show with six contours 
the cosmic ray energy density $e_c(r,z)$ 
in units of $10^{-12}$ erg cm$^{-3}$.
Two adjacent contours are labeled and others can be found 
by extending the same additive variation.
}
\label{f1}
\end{figure}

\clearpage
\begin{figure}%2
\vskip.5in
\centering
\hspace{2.in}
\includegraphics[bb=62   5 350 437,scale=1.0,angle= 0]
{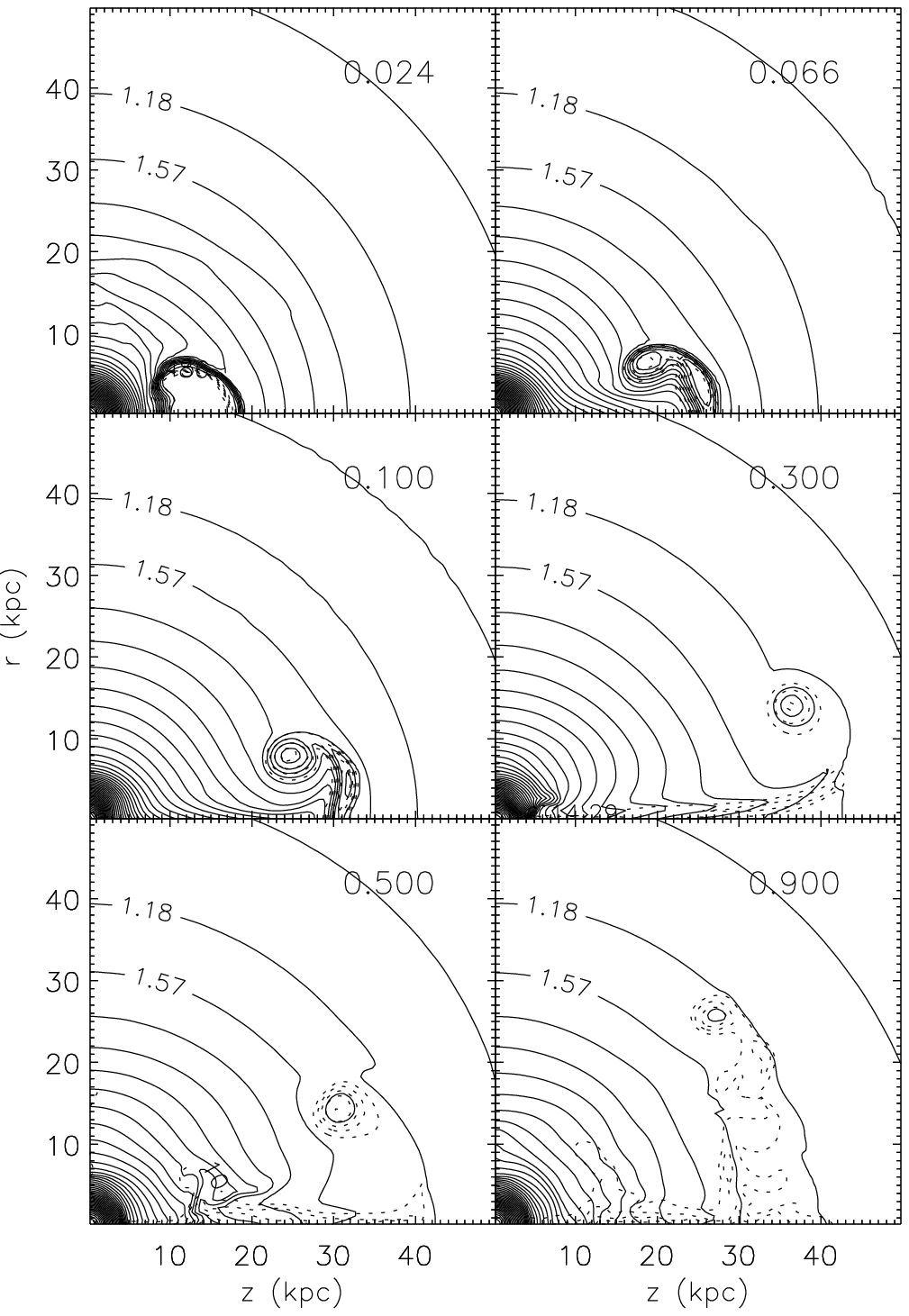}
\vskip.5in
\caption{
Evolution of an X-ray cavity in Virgo with cosmic ray diffusion
parameter $n_{e0} = 6\times 10^{-6}$ cm$^{-3}$
at six times shown in Gyrs at the upper right of each panel.
Solid lines show the gas density contours $\rho(r,z)$ in units of
$10^{-26}$ g cm$^{-3}$.
Dotted lines show with six contours 
the cosmic ray energy density $e_c(r,z)$
in units of $10^{-12}$ erg cm$^{-3}$.
Two adjacent contours are labeled and others can be found
by extending the same additive variation.
Due to crowding at early times 
the cosmic ray contours are difficult to distinguish 
in the first three panels.
For the final three panels the outer two cosmic ray contours are 
14.29 \& 31.43 (panel 0.300), 
10.0 \& 20.0 (panel 0.500), and 
4.29 \& 11.43 (panel 0.900).
}
\label{f2}
\end{figure}

% new post-referee figure:
\clearpage
\begin{figure}%3
\vskip1.5in
\centering
\includegraphics[bb=90 166 522 519,scale=0.8,angle= 0]
{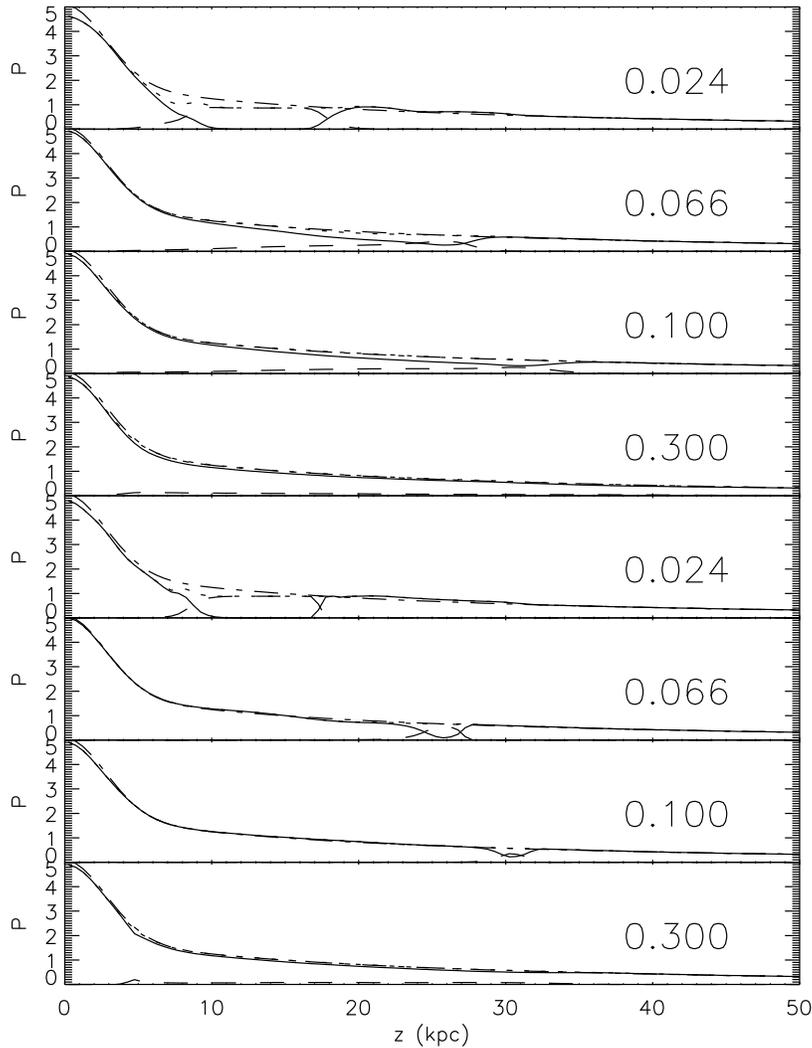}
\vskip1.7in
\caption{
Pressure profiles along the $z$-axis in units of 
$10^{-10}$ dynes cm$^{-2}$. 
Each panel contains four (often overlapping) profiles: 
the gas pressure $P(r)$ (solid lines), cosmic ray pressure 
$P_c(r)$ (long dashed lines), the total pressure 
$P + P_c$ (dotted lines), and 
the initial gas pressure in the cluster before the cavity 
is introduced (dash-dotted lines).
{\it Upper four panels:} show pressure profiles 
at four times for the 
evolution in Fig. 1 with $n_{e0} = 6\times 10^{-3}$ cm$^{-3}$.
{\it Lower four panels:} show pressure profiles
at four times for the
evolution in Fig. 2 with $n_{e0} = 6\times 10^{-6}$ cm$^{-3}$.
Each panel is labeled with the time in Gyrs.
}
\label{f3}
\end{figure}

\clearpage
\begin{figure}%4
\vskip2.5in
\centering
\includegraphics[bb=90 166 522 519,scale=0.8,angle= 0]
{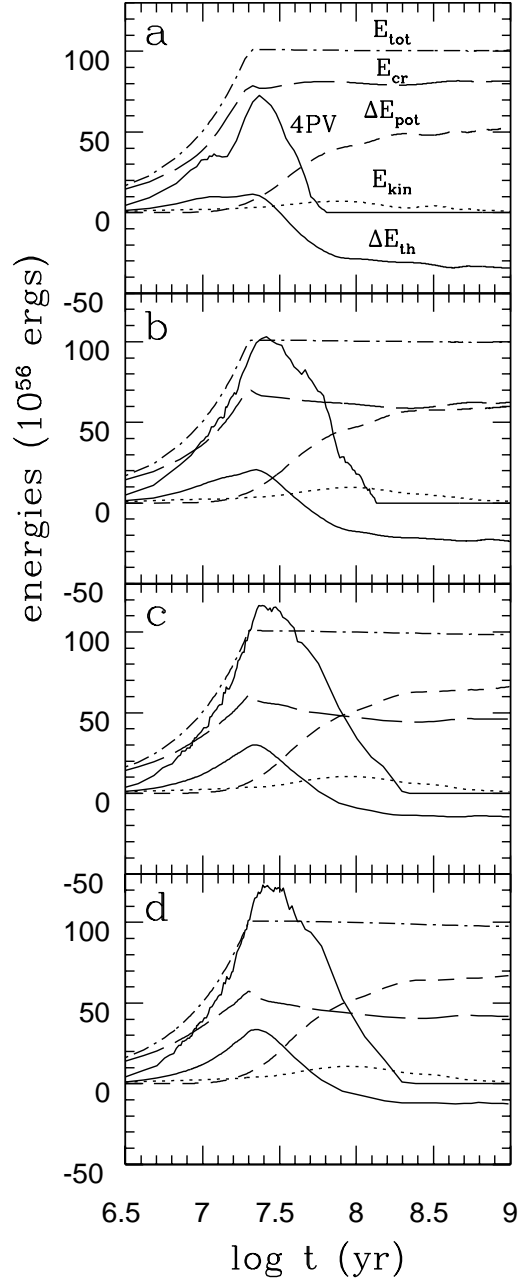}
\vskip1.7in
\caption{
Global energy evolution in four cavities
with decreasing cosmic ray diffusion coefficient
$\kappa \propto n_{e0}$ characterized by density parameters 
$n_{e0}$: 
$6\times 10^{-3}$ (panel $a$),
$6\times 10^{-4}$ (panel $b$),
$6\times 10^{-5}$ (panel $c$), and
$6\times 10^{-6}$ (panel $d$).
The energies are labeled as follows: 
cosmic ray energy $E_{cr}$ (long dashed lines);
change in potential energy $\Delta E_{pot}$ (short dashed lines);
kinetic energy $E_{kin}$ (dotted lines);
change in thermal energy $\Delta E_{th}$ (lower solid lines);
and the total energy $E_{tot}$ (dash-dotted lines).
The total energy associated with the approximate cavity volume 
$4PV$ is shown in the upper solid lines.
All energies are in units of $10^{56}$ ergs and are those in 
the hemisphere containing our computational grid.
}
\label{f4}
\end{figure}

\clearpage
\begin{figure}%5
\vskip1.5in
\centering
\hspace{-1in}
\includegraphics[bb=90 166 522 519,scale=0.8,angle= 270]
{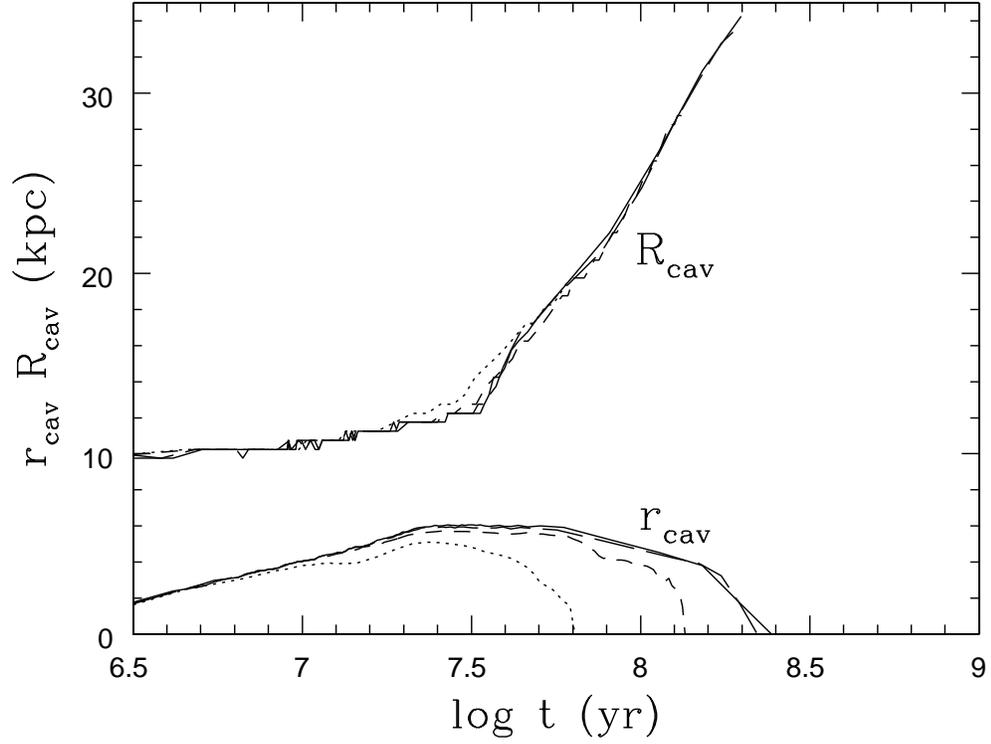}
\vskip.7in
\caption{
Approximate evolution of the cavity radius
$r_{cav}(t)$ and its mean radius in the cluster $R_{cav}(t)$.
The cavity radius is found from the estimated cavity
volume $V$ by assuming that the cavity is spherical,
$r_{cav} = [3V/4\pi]^{1/3}$.
The four lines correspond to the four 
decreasing cosmic ray diffusivities
$\kappa(n_e,n_{e0})$ in each panel of Figure 4:
$a$, dotted line;
$b$, short dashed line;
$c$, long dashed line;
$d$, solid line.
Each overlapping trajectory $R_{cav}(t)$ ends at the 
time when $r_{cav} \rightarrow 0$.
}
\label{f5}
\end{figure}

\clearpage
\begin{figure}%6
\vskip2.in
\centering
\hspace{-1in}
\includegraphics[bb=90 166 522 519,scale=0.8,angle= 270]
{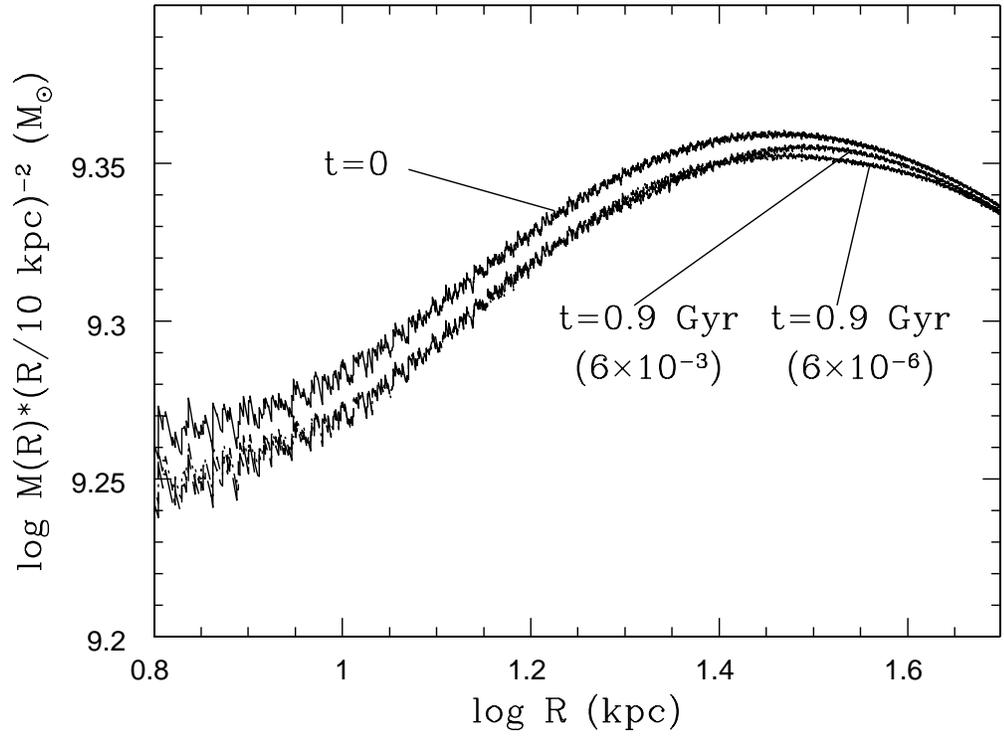}
\vskip.7in
\caption{
Variation of the cumulative spherical mass gas distribution
$M(R)$ in one hemisphere of 
the cluster at times $t = 0$ and 0.9 Gyr for
two values of the cosmic ray diffusion parameter $n_{e0}$ 
shown in parentheses.
The cumulative mass is multiplied by $(R/10~{\rm kpc})^{-2}$
to remove most of the radial variation between $R = 6.3$
and 50 kpc, the approximate Virgo cooling radius.
}
\label{f6}
\end{figure}

\clearpage
\begin{figure}%7
\centering
\vskip0.5in
\hspace{2.in}
\includegraphics[bb=62   5 350 437,scale=1.0,angle= 0]
{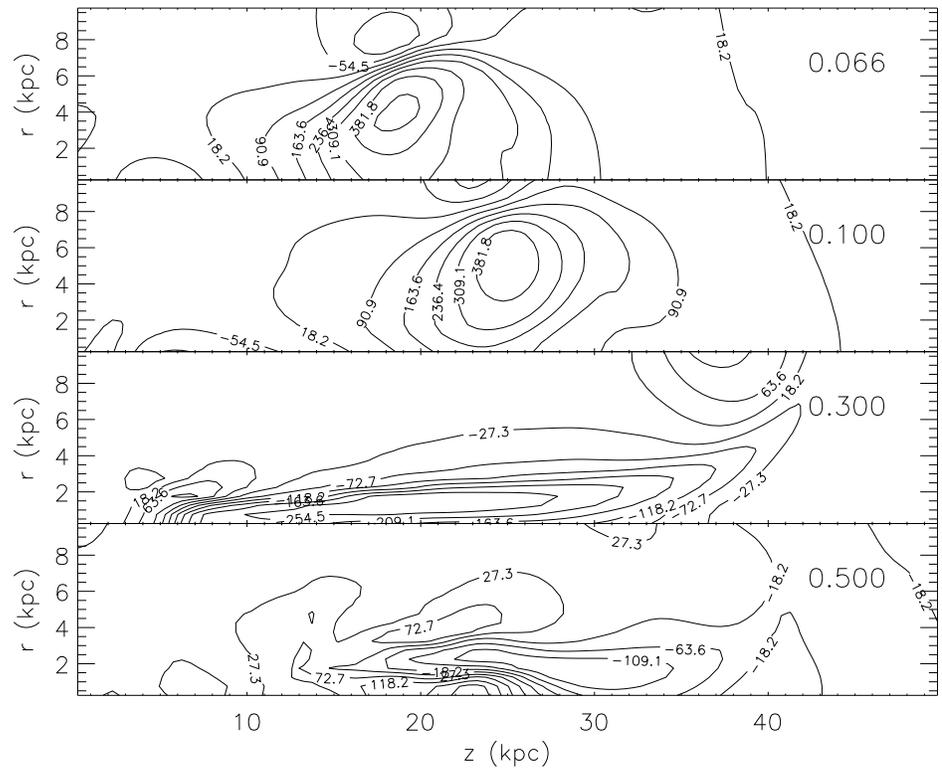}
\vskip.5in
\caption{
Evolution of the radial velocity $v_{r}(r,z)$ near the $z$-axis 
with cosmic ray diffusion
parameter $n_{e0} = 6\times 10^{-6}$ cm$^{-3}$
at four times shown in Gyrs at the upper right of each panel.
Contours are labeled with values of $v_{r}$ in km s$^{-1}$.
}
\label{f7}
\end{figure}

\clearpage
\begin{figure}%8
\centering
\vskip1.5in
\includegraphics[bb=90 166 522 519,scale=0.8,angle= 270]
{f8.eps}
\vskip1.in
\caption{
{\it Top panel}: Correlation of the cosmic ray partial pressure 
$P_c/(P + P_c)$ 
with the entropy change $\Delta s$ 
in each computational zone for the low-$\kappa$ 
post-cavity flow over time $t = 0.9$ Gyrs.
Each point represents a computational zone within 
cluster radius $R = 50$ kpc.
{\it Bottom panel}: Location of regions in the cluster with 
high cosmic ray partial pressures ($P_c/(P + Pc) > 0.3$) (points) 
and regions of large negative entropy change 
($\Delta s < -0.1$) (open squares). 
}
\label{f8}
\end{figure}

\end{document}